\documentclass[lettersize,journal]{IEEEtran}

\usepackage{amsmath,amsfonts}
\usepackage{algorithm}
\usepackage{array}
\usepackage{textcomp}
\usepackage{stfloats}
\usepackage{url}
\usepackage{verbatim}
\usepackage{graphicx}
\usepackage{cite}
\usepackage[colorlinks]{hyperref}
\usepackage{paralist}

\usepackage[utf8]{inputenc}
\usepackage{amsmath}
\usepackage{multirow}
\usepackage{multicol}
\usepackage{booktabs}
\usepackage[flushleft]{threeparttable}
\usepackage{xspace}
\usepackage{caption}
\usepackage{graphicx}
\usepackage{url}
\usepackage{algorithm}
\usepackage[noend]{algpseudocode}
\usepackage{amssymb}
\usepackage{dsfont}
\usepackage{subfigure}
\usepackage{makecell}
\usepackage{tablefootnote}
\usepackage{xcolor}
\usepackage{multirow}
\newcommand\blfootnote[1]{%
  \begingroup
  \renewcommand\thefootnote{}\footnote{#1}%
  \addtocounter{footnote}{-1}%
  \endgroup
}

\newcommand{\name}{CableMon\xspace} % Monitor for the cable network
\newcommand{\sys}{CableMon\xspace} % Monitor for the cable network
\newcommand{\Name}{CableMon\xspace}
\newcommand{\precision}{ticket prediction accuracy\xspace}
\newcommand{\recall}{ticket coverage\xspace}

\newcommand{\pnmdata}{PNM data\xspace}

\newcommand{\FN}{FN\xspace}

\newcommand{\orangedots}{dots\xspace}
\newcommand{\reddot}{square\xspace}
\newcommand{\reddots}{squares\xspace}

\newcommand{\greendots}{triangles\xspace}

\newcommand{\anonisp}{AnonISP\xspace}

\newcommand{\PENDING}[1]{#1}

\newcommand{\ie}{{i.e.}\xspace}
\newcommand{\ea}{{et al.}\xspace}

\newcommand{\xy}[1]{#1}

\newcommand{\zcut}[1]{}

\expandafter\def\expandafter\normalsize\expandafter{%
    \normalsize%
    \setlength\abovedisplayskip{0pt}%
    \setlength\belowdisplayskip{0pt}%
    \setlength\abovedisplayshortskip{-8pt}%
    \setlength\belowdisplayshortskip{2pt}%
}

\begin{document}

\title{Improving the Reliability of Cable Broadband Networks
  via Proactive Network Maintenance}

\author{Jiyao Hu\hyperlink{author}{$^*$}, Zhenyu Zhou\hyperlink{author}{$^*$}, Xiaowei Yang,~\IEEEmembership{Duke University}
        % <-this % stops a space
}



\maketitle

\begin{abstract}
   
Cable broadband networks are one of the few ``last-mile'' broadband
technologies widely available in the U.S. Unfortunately, they have
poor reliability after decades of deployment. The cable industry
proposed a framework called Proactive Network Maintenance (PNM) to
diagnose the cable networks. However, there is little public knowledge
or systematic study on how to use these data to detect and localize
cable network problems. Existing tools in the public domain have
prohibitive high false-positive rates.  In this paper, we propose
\name, the first public-domain system that applies machine learning
techniques to PNM data to improve the reliability of cable broadband
networks. \xy{\name tackles two key challenges faced by cable ISPs:
  accurately detecting failures, and distinguishing
  whether a failure occurs within a network or at a subscriber's
  premise. \Name uses statistical models to generate features from
  time series data and uses customer trouble tickets as hints to infer
  abnormal/failure thresholds for these generated features. Further,
  \name employs an unsupervised learning model to group cable devices
  sharing similar anomalous patterns and effectively identify
  impairments that occur inside a cable network and impairments occur
  at a subscriber’s
  premise, as these two different faults require different types of
  technical personnel to repair them.}  We use eight months of PNM
data and customer trouble tickets from an ISP and experimental deployment to evaluate \name's
performance. Our evaluation results show that \name can effectively detect and distinguish failures from PNM data and outperforms existing public-domain tools. 
    
    \blfootnote{\hypertarget{author}{$^*$~Jiyao Hu and Zhenyu Zhou, placed in alphabetic order, contributed equally to this work.}}
\end{abstract}


\section{Introduction}
\label{sec:intro}

Broadband access networks play a crucial role in modern life. They
help narrow the digital divide, enable e-commerce, and provide
opportunities for remote work, study, and entertainment.  In the US,
cable networks are one of the few available infrastructures that can
provide broadband Internet access to US homes. In many rural areas,
they are often the only broadband choice.  According to a study in
2016~\cite{cablelabs2016broadband}, cable broadband is available to
93\% of US homes, far more than the two alternative choices: Very-high-bitrate Digital Subscriber Line (VDSL) (43\%) and
Fiber-to-the-Premises (FTTP) (29\%).

However, cable networks are prone to failures, partly due
to the nature of their Hybrid Fiber-Coaxial (HFC) architecture. This architecture
uses both optical fibers and coaxial cables to deliver a mixed bundle
of traffic, including video, voice, and Internet data. Unlike fiber
optics, coaxial cables are vulnerable to radio frequency (RF)
interference.  Many parts of the cable networks are now decades
old~\cite{historyofcable}. Aging can lead to problems such as cable
shielding erosion, loose connectors, and broken amplifiers. All those
problems can manifest themselves as poor application layer
performance, e.g., slow web responses or low-quality video
streaming. Many measurement studies have shown that broadband networks
have poor
reliability~\cite{grover2013peeking,padmanabhan2019residential,bischof2017characterizing,lehr2011assessing,tom2015intermittent,larry2016comcast}.
A recent one~\cite{bischof2017characterizing} shows that the average
availability of broadband Internet access is at most two nines (99\%),
much less than the minimum Federal Communications Commission (FCC)'s 

requirement (four nines 99.99\%) for
the public switched telephone networks (PSTNs)~\cite{kuhn1997sources}.
Admittedly, if ISPs replace the last-mile coaxial cables with fiber
optics, many of these problems may disappear. However, due to the
the prohibitive cost of FTTP, cable broadband networks are likely to
remain as one of the few broadband choices in rural America for the
next decade or two. Therefore, it is critically important that 
cable Internet services remain robust during emergencies, especially
as more and more subscribers migrate their landline phones to VoIP
phones.

The cable industry has long recognized this problem and developed a
platform called Proactive Network Maintenance (PNM) to improve the
reliability of their networks~\cite{cablelabs2016docsis3}. PNM enables
a cable ISP to collect a set of performance metrics from each
customer's cable modem.  We refer to this set of data as PNM data. One
example of a PNM metric is a cable channel's signal-to-noise
ratio. PNM aims to enable an ISP to proactively detect and fix network
faults before they impact services and customers.

Although PNM has been incorporated into DOCSIS since
2005~\cite{cablelabs2016docsis3}, how to use PNM data to improve
network reliability remains an open challenge. \xy{Ideally, an operator should be able to use the PNM data collected from their networks to quickly detect failures, and distinguish the types of failures so as to dispatch the right repair team.} However, 
the best current
practice recommended by CableLabs\footnote{CableLabs is a research and
  development lab founded by American Cable operators in 1988 and led
  the development of DOCSIS and PNM.}~\cite{cablelabs2016docsis3} and
the tools used by some ISPs~\cite{larry2016comcast} use a set of
manually configured thresholds to flag a faulty network condition. The
feedback from deploying ISPs is that these thresholds are often too
conservative, leading to high false positives. \xy{In addition, there is a lack of  publicly available tools that use the PNM data to distinguish common types of failures in the cable broadband infrastructure. ISPs often rely on manual methods to diagnose failure types, a  process prone  to errors and frequently results in dispatching the wrong type of repair personnel.} 

This work aims to improve the reliability of cable broadband
networks. \xy{To this end, we develop a system called \name to assist
  a cable ISP in detecting and diagnosing failures.} \Name includes
two central components. The first component is a PNM-based fault
detection system using PNM data. A main challenge faced by the design of \name is the lack of expert
knowledge and ground truth on what PNM values warrant a proactive
network repair. In an RF system like a cable network, network
conditions may degrade gradually, making it difficult to define a
static threshold that separates what is faulty from what is not.  To
address this challenge, we use
machine learning techniques to infer network faults that demand
immediate repair. \Name couples PNM data with customer trouble tickets
to identify the ranges of PNM values that are likely to lead to a
customer's trouble call. We hypothesize that if a network fault
impacts an ISP's customers, then some customers are likely to call to
report the problem. Therefore, we can use customer trouble tickets as
hints to learn what network conditions are likely to lead to customer
trouble calls.  An ISP should prioritize its efforts to repair those
problems, because if they persist, they are likely to reduce customer
satisfaction and increase the cost of customer support.

\xy{At the heart of the second component of \name is a clustering algorithm that distinguishes failures occurred inside a cable broadband network from those occurred inside a subscriber's premise, as these two types of faults require an ISP to dispatch different repair teams (\S~\ref{sec:types-of-faults}). Accurate diagnosis of network faults from subscriber-premise faults can reduce dispatch errors, leading to shortened failure repairment times and operational cost savings.}

With the support of CableLabs, we have obtained eight months of
anonymized PNM data and the corresponding customer trouble tickets
from a mid-size U.S. ISP. We use five months of data to train \name,
and use the next three months' data following the training set as the
test set to evaluate how well \name detects network faults. \name
takes the PNM data in our test set as input and detects when a network
fault occurs and when it ends. Due to the lack of ground truth, we
evaluate \name's performance using customer trouble tickets in the
test set. When \name detects a network anomaly, if a customer who
experiences the anomaly reports a ticket, we consider the detection
a success. We compare \name\xy{'s fault detection effectiveness} with a tool currently used by our
collaborating ISP, which we refer to as AnonISP, and with a tool
developed by Comcast~\cite{larry2016comcast}. Our results show that
$81.9\%$ of the anomalies\footnote{In this work, we use the words 'failures', 'faults', and 'anomalies' interchangeably.} 
detected by \name lead to a customer trouble
ticket. In contrast, only $10.0\%$ of the anomalies detected by
AnonISP's tool lead to a trouble ticket; and $23.5\%$ of the anomalies
detected by Comcast's tool lead to a customer ticket. In addition,
\name predicts $23.0\%$ of all network-related tickets, while
AnonISP's tool predicts $25.3\%$ and Comcast's tool predicts less than
3\%. The trouble tickets predicted by \name on average last 32.5 hours
(or 53.3\%) longer than those predicted by other tools, suggesting
that those tickets are more likely to require repair efforts.  The
median time from the beginning of a fault detected by \name to the
reporting time of a ticket is 164.1 hours (or 29.3\%) shorter than
that of a fault detected by other tools, suggesting that the faults
detected by \name require more immediate repair.

\xy{Furthermore, we evaluate how effectively \name separates network faults from subscriber-premise faults. Using the anonymized PNM data and the corresponding customer trouble tickets
from the same U.S. ISP, we
manually labeled a small set of devices as healthy, experiencing a
network failure, or experiencing a subscriber-premise failure. 
Using the
manually labeled data as ground truth, we show that \sys's clustering
algorithm achieves a rand index~\cite{rand1971objective} of 0.91 (1.0
being the highest), indicating that \sys's fault clustering is highly
accurate.}  \PENDING{\anonisp also confirms the effectiveness of \sys's fault categorization with practical field tests.}

To the best of our knowledge, this work is the first large-scale
public study that couples PNM data with customer trouble tickets to
improve the reliability of cable networks. Our main contribution is
\name, a system that detects network faults more reliably than the
existing public-domain work\PENDING{, and also the first system that automatically distinguishes network faults from subscriber-premise faults}. It serves as a starting point to unleash
the full potential of PNM data. One general
lesson we learn is that one can use customer trouble tickets as hints
to learn what values of network performance metrics indicate
customer-impacting problems, despite the presence of noise in both the
ticket data and the network performance data. We believe this lesson
is applicable to proactive network maintenance in other types of
networks, including cellular networks, WIFI access networks, and
datacenter networks.

\section{Background and Datasets}

In this section, we briefly introduce the cable Internet
architecture and describe the datasets we use in this work.

\subsection{Cable Network Architecture}

\begin{figure}[tb!]
\centering
    \includegraphics[width=.5\textwidth]{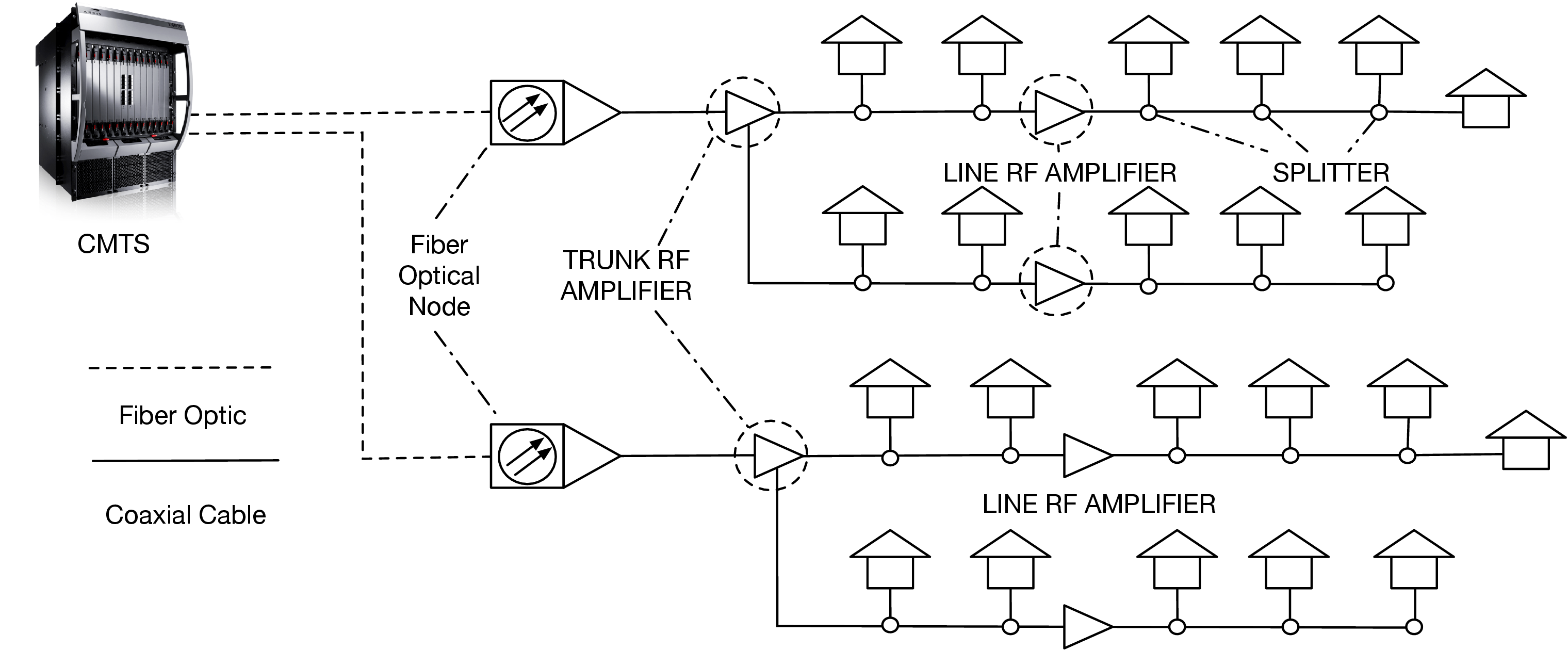}
    \caption{\small\textbf{An overview of the Hybrid Fiber Coaxial (HFC) architecture.}}
    \label{f:hfc}
\end{figure}

Figure~\ref{f:hfc} shows a high-level overview of a cable broadband
network. A cable broadband network is an access network. It provides
the ``last-mile'' Internet connectivity to end users. A customer
connects to the cable network via a cable modem residing in her
home. The cable access network terminates at a device called a Cable
Modem Termination System (CMTS), which is a router with one port
connecting to the Internet and many other ports connecting to
customers' cable modems.

At the IP level, there is only one hop between a customer's cable
modem/home router and the CMTS. Underlying this single IP hop, there
is a complicated link-level structure that consists of different types
of physical links and devices.  The ``last-mile'' links that connect
to the customer premises are often made of copper coaxial cables.
These cables terminate at a device called a fiber node (FN). A fiber
node connects to the CMTS via optical fibers.  It converts the
incoming optically modulated signals into electrical signals and sends
the signals toward the customers' homes, and vice versa.  Due to
signal attenuation, cable networks deploy radio frequency (RF)
amplifiers between a fiber node and a residential home. Along the way
to a customer's home, new branches may split from the main cable by
the line splitters. All these devices could introduce signal
distortion and noise.

Historically, cable TV networks divide radio frequency into multiple
channels, each of 6MHz width. Cable broadband networks use a subset of
these channels as data carriers. A cable ISP typically uses three or
four of these channels at the lower end of the spectrum to carry data
from a user's cable modem to CMTS. We refer to this direction as the
\textit{upstream} direction. An ISP may use sixteen or more of the
channels at the higher end of the spectrum to carry data from a CMTS
to a modem. We refer to this direction as the \textit{downstream}
direction.

\subsection{Types of Faults}
\label{sec:types-of-faults}

\PENDING{There are two types of common faults that impact the service quality
and availability of cable broadband networks.  The first type of fault
is a maintenance issue, where a faulty component lies inside the customer-shared 
network infrastructure.  The second type of fault is a service issue, where a
faulty component lies in a subscriber's premise.  Distinguishing a maintenance issue from a service issue among the devices with anomalies is important, because repairing
each type of fault requires a different type of technician. If a cable
ISP makes a wrong diagnosis, they may send a service technician to a
subscriber's home for a maintenance issue or vice versa. In such
cases, the technician is unable to repair the fault, resulting in a waste of 
operational resources and a delay in failure repair time. 
}

\subsection{Datasets}

We have obtained two types of anonymized modem-level data from a U.S. cable
ISP for this study. They include (1) PNM data and (2) customer trouble
ticket data.  We have a total of eight months of data dating from 01/06/2019
to 08/31/2019.  Next, we describe each dataset in turn.\footnote{We note
  that we have discussed this work
with our institute's IRB. And they consider it does not involve
human subjects.}

\noindent \textbf{PNM data:} The PNM data we obtained were collected
by a common standard built into DOCSIS. A CMTS can query a
DOCSIS-compliant cable modem (CM) to obtain certain
performance data. DOCSIS standardizes how a CM or CMTS stores these
performance data in a local Management Information Base
(MIB)~\cite{cablelabs2016docsis3}. A remote process can use the Simple
Network Management Protocol (SNMP) to query the MIBs of each CM or a
CMTS to obtain performance data~\cite{rfc4036}.

Currently, we only have PNM data from the upstream channels. DOCSIS
3.0 gives a cable operator the ability to collect the full spectrum
view of a cable modem's RF channels. It is our future work to
investigate whether this type of data may further improve our
detection accuracy. 

A record in the PNM data we obtain has the following fields:
\begin{compactitem}
\item \emph{Timestamp}: The time when a PNM query is sent.
\item \emph{Anonymized MAC}: The hashed MAC address of the queried CM.
\item \emph{Anonymized Account Number}: The hashed user account number. This
  field is used to link a customer ticket with the corresponding PNM
  data from the customer's CM.
\item \emph{Channel Frequency}: This field identifies which upstream channel
  this record is about. 
\item \emph{SNR}: The upstream signal-to-noise ratio of this channel.
\item \emph{Tx Power}: A CM's signal transmission power.
\item \emph{Rx Power}: The received signal power at the CMTS.
\item \emph{Unerrored}: The number of unerrored codewords received at the
  CMTS.
\item \emph{Corrected}: The number of errored but corrected codewords
  received at the CMTS.
\item \emph{Uncorrectable}: The number of errored but uncorrected codewords.

\item \emph{T3 Timeouts}: The number of DOCSIS T3
  timeouts~\cite{docsis30} the CM has experienced since its last
  reboot. A DOCSIS T3 timeout occurs when there is an error in a CM's
  ranging process, which we will soon explain.
\item \emph{T4 Timeouts}: The number of DOCSIS T4
  timeouts~\cite{docsis30} the CM has experienced since its last
  reboot. Similarly, a T4 timeout occurs when there is a ranging
  error. 
\item \emph{Pre-Equalization Coefficients}: The set of parameters a CM uses
  to compute how to compensate for channel distortions during a
  ranging process.
\end{compactitem}

A CM uses a process called \textit{ranging} to compute a set of
parameters called \textit{pre-equalization coefficients} for
mitigating channel distortions.  When RF signals travel along a
coaxial cable, they may be distorted as different frequencies
attenuate at different speeds and noise may be added to the
channel. To mitigate the channel distortions, a CM adds correction
signals to the data signals it transmits.  Ideally, the correction
signals will cancel out the distortions when the signals arrive at the
CMTS.  A CM and a CMTS exchange messages periodically to compute the
correction signals. This process is called ranging. And the set of
parameters used to compute the correction signals are called
pre-equalization coefficients.

The PNM data we obtain are collected every four hours from several of
an ISP's regional markets. There are around 60K unique account numbers
in our datasets.

\noindent\textbf{Customer Ticket Data:} We have also obtained the
records of customer trouble tickets from the same ISP.  The relevant
fields in each record include the hashed customer's account number,
the ticket creation time, the ticket close time (if it was closed), a
brief description of the actions taken to resolve the ticket, a
possible diagnosis, \PENDING{and a category of the
issue based on the ISP's diagnosis. The category includes two
classes: a part-of-primary ticket or not. The last field is crucial 
to separate network faults from isolated faults occurring in a subscriber's premise. A part-of-primary ticket indicates that the ISP considers
the issues the customers are experiencing a maintenance issue. Thus,
it groups the tickets as one conceptual ``primary'' ticket. All
part-of-primary tickets that belong to the same maintenance issue have
the same primary ticket identifier. In this work, we refer to
part-of-primary tickets as \emph{maintenance tickets} and other
infrastructure-related tickets as \emph{service tickets}.}

\begin{table*}[!t]
\begin{center}
\begin{tabular}{c|c|c|c|c|c|c}
& 03/13/2019 & 04/09/19 & 06/25/19 & 07/15/19 & 08/15/19 & Eight-month  \\
 \midrule
 MTR $<$ 18dB & 24.95 \% & 25.45 \% & 27.16 \%  & 27.07 \% & 27.38 \% & 26.15 \% \\
\end{tabular}
\end{center}
\caption{\small\textbf{The percentage of cable modems that need to be
    repaired if an ISP were to follow one of the CableLabs'
    recommendations. \label{tab:mtr}}}
\end{table*}

\section{Overview}

In this section, we motivate the design of \name by describing the
limitations of existing work. We then describe the design rationale of
\name, its design goals, and the design challenges we face.

\subsection{Limitations of Existing Work}

The existing PNM work in the public domain~\cite{cablelabs2016docsis3,
  larry2016comcast,tom2015intermittent} use a set of PNM metrics and
manually-set thresholds to detect network faults. If the value of a
metric is below or above a threshold, it indicates a fault.  This
approach has several limitations. First, it is challenging to set the
right thresholds. If the thresholds were set too conservatively, they
might flag too many faults for an ISP to fix. In contrast, if they
were set too aggressively, an ISP might miss the opportunities for
proactive maintenance. There lacks a systematic study on how to set
the threshold values to achieve the best tradeoff.  Second, the
existing work mostly uses the instantaneous values of PNM data for
fault detection. However, due to the inherent noise in PNM data, using
the instantaneous values may lead to instability in detection
results. In addition, it may fail to detect faults that can only be
captured by abnormalities in a PNM metric's statistical values, e.g.,
variations. \PENDING{Finally, they cannot tell if a network fault is a service or maintenance issue.}

For ease of explanation, we use one threshold value recommended in the
CableLabs' PNM best practice document~\cite{cablelabs2016docsis3} to
illustrate the limitations. CableLabs' recommendation uses a variable
called Main Tap Ratio (MTR) computed from a modem's pre-equalization
coefficients. It specifies that when the MTR value of a modem is below
a threshold ($<$18dB), there is a fault in the network that needs
immediate repair.

We sample the MTR values in one of the ISP's markets. There are more
than 60K modems in this market.  We choose five random days' records
during an eight-month period in 2019 and measure the MTR values of all
modems during the sampled days. Table~\ref{tab:mtr} shows the
percentage of modems that have a channel whose MTR value is below the
recommended threshold. If an ISP used the recommended MTR threshold,
at any sampled day, there would be more than 24\% of cable modems that
require immediate repair.  We also measure the MTR values among all
PNM records during this eight-month period.  In more than 26\% of the
records, a modem's MTR value is below 18dB.

\subsection{Design Goals}

\name aims to enable an ISP to detect network problems that demand
immediate repair\PENDING{, and deliver proper repair}.  Specifically, it aims to accurately detect the set
of network conditions that adversely impact customer experience\PENDING{, and whether they are service or maintenance issues}. We
refer to these network conditions as network anomalies or faults in
this work. Its design goals include the following:

\begin{itemize}

\item High ticket prediction accuracy, and moderate ticket
  coverage. Ideally, we would like to use precision (the set of true
  positives detected over all detected positives) and recall (the set
  of true positives detected over all true positives) to measure the
  performance of \name. However, because we do not know the ground truth,
  we instead use customer tickets as indications of true positives.
  We define \textit{\precision} as the ratio between the number of
  anomalies detected by \name that lead to one or more customer
  tickets and the number of total anomalies \name detects.  Similarly,
  we define \textit{\recall} as the ratio between the number of
  tickets \name predicts and the total number of network-related
  customer tickets.  It is desirable that \name has high ticket
  prediction accuracy because an ISP is often limited by the number of
  technicians it has to repair network faults, avoiding false alarms
  is practically more important than repairing all faults
  proactively. What we learned from AnonISP is that even a 10\%
  reduction in customer calls can reduce their operational costs
  significantly. Therefore, as a starting point, we aim for a high
  \precision and a moderate \recall.

\item No manual labeling. One approach to detect network anomalies is
  to train a supervised learning classifier on labeled data. The
  labels tell what PNM metrics indicate network anomalies and what do
  not. However, we do not have such labeled data. And due to the lack of
  ground truth and the large size of the data, manual labeling is also
  practically challenging. Therefore, we aim to design \name without
  requiring manual labeling.

\item No extensive parameter tuning. We aim to release \name as an
  off-the-shelf-tool at cable ISPs. Therefore, we require that \name's
  fault detection methods work effectively without much parameter
  tuning on the ISP side.

\item Efficient. We require that \name can detect whether there is a
  network fault or not in real time. This is because an ISP can deploy
  \name as a diagnosis tool in addition to using it for proactive
  network maintenance.  When an ISP receives a customer trouble call,
  it is often challenging to diagnose what has caused the customer's
  problem. An ISP can use \name to help diagnose whether the problem
  is caused by a network fault.

\end{itemize}

\begin{figure}[tb!]
\centering
    \includegraphics[width=.5\textwidth]{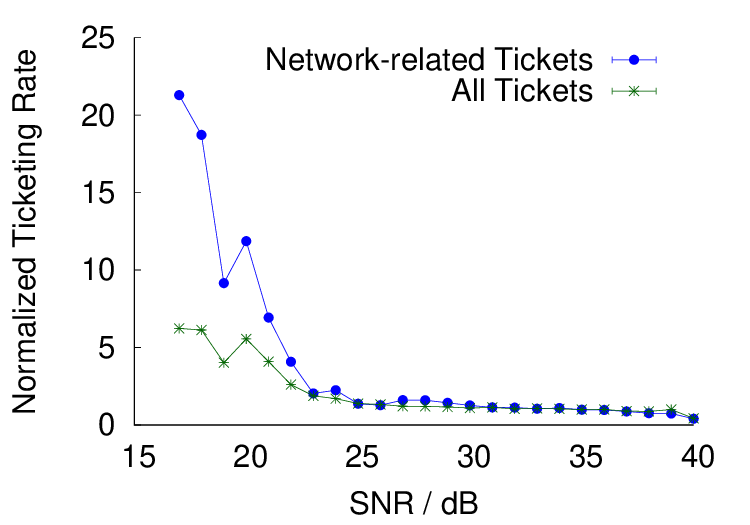}
    \caption{\small\textbf{This figure shows how the customer
        ticketing rate varies with the values of SNR. Ticketing
        rate tends to increase when SNR values are low.}}
    \label{f:snr_rate}
\end{figure}

\begin{figure*}[t!]
% \captionsetup[subfigure]{aboveskip=-1000pt,belowskip=-1000pt, skip=-1000pt}
\centering \subfigure[]{
  \includegraphics[width=.4\textwidth]{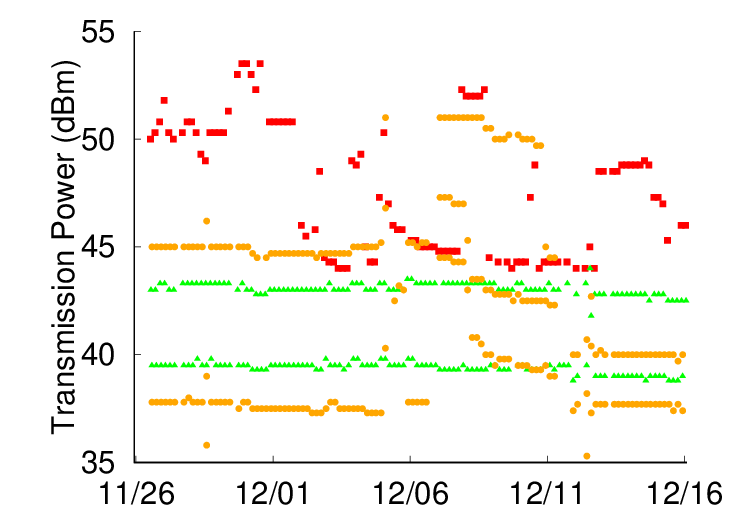}
    \label{f:tx_power_examples}
    } \hspace{10pt}
    \subfigure[]{
    \includegraphics[width=.4\textwidth]{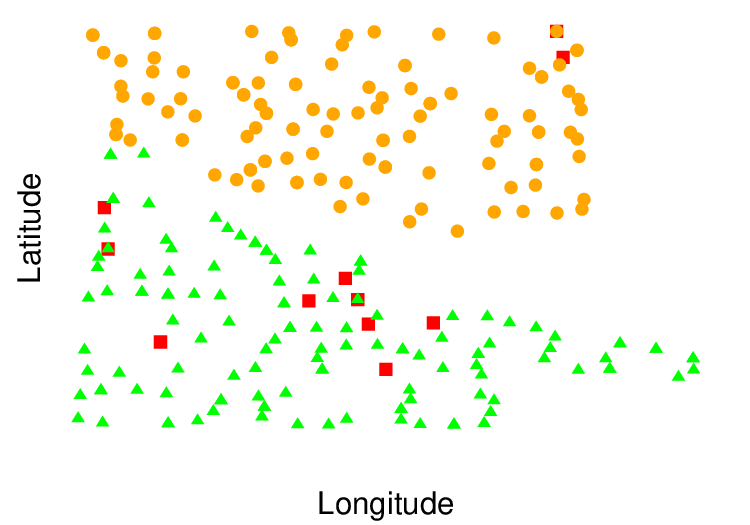}
    \label{f:tx_power_examples_map}
    }
    \vspace{-10pt}
    \caption{{\small{\bf Figure~(a) shows how the transmission powers of several 
    cable devices in the same fiber optical node fluctuate over time. Orange \orangedots are devices that show the same anomalous transmission power patterns. Green \greendots are devices that show normal patterns. Red \reddots are devices that show distinct anomalous patterns. Figure~(b) shows the locations of the cable devices using the same colored icons.}}\label{f:align}}
\end{figure*}

\subsection{Design Rationale}\label{sec:design_rationale}

To meet \name's design goals, we use customer trouble tickets as hints
to train a customized classifier to detect network faults. We
hypothesize that the occurrences of customer trouble tickets should
correlate with those of network faults. When a customer-impacting
fault occurs, some customers are likely to call the ISP to fix it. Each
call creates a trouble ticket. If the values of PNM data can
indicate network faults, then the values of PNM data should
correlate with how frequently customer trouble tickets are created. In
this paper, we define the average number of customer tickets created
in a unit time as the \emph{ticketing rate}.

To validate this hypothesis, we measure how ticketing rate changes
with different values of a PNM metric. For a PNM metric $m$ (e.g.,
SNR), we sort the values of $m$ in an ascending order. Each pair of
adjacent values defines a bin $b$. For each bin $b$, we measure the
number of tickets $N_b$ that occur in the time periods where the value
of $m$ falls within the bin, and the total length of those time
periods $T_b$.  We then divide $N_b$ by $T_b$ to obtain the ticketing
rate for bin $b$.  We note that a PNM record is collected at discrete
time points (roughly four hours apart in our datasets). We assume that
a PNM value remains unchanged between its collection points.

As an example, we show how the ticketing rate varies with the values
of SNR in Figure~\ref{f:snr_rate}. We normalize this value by the
baseline ticketing rate, which we obtain by dividing the total number
of customer tickets in our dataset by the total collection period. The
line marked by the legend ``All Tickets'' shows how the ticketing rate
varies with the values of SNR if we consider all tickets; and the line
marked by ``Network-related Tickets'' shows how the ticketing rate of
network-related tickets varies with SNR. As can be seen, when the
values of SNR are low, both the network-related ticketing rate and the
all-ticket ticketing rate tends to increase, suggesting that low SNR
values signal network faults.

In practice, customer tickets do not always indicate network
faults. On the one hand, many customers may call an ISP for
non-network related problems. The customer ticket data we obtain
includes a ticket action field and a ticket description field, which
provide the information on how an ISP deals with a ticket.  We observe
that nearly 25\% of tickets are resolved via ``Customer Education'' or
``Cancelled'', suggesting that they are not caused by networking
problems. On the other hand, customers may not report tickets when
network outages indeed take place.  In our ticket dataset, when an
outage affects an entire region, all tickets caused by that outage are
labeled as ``part of primary'', grouped and pointed to a primary
ticket, which is a representative ticket of the outage.  We manually
checked an outage that affected more than 200 customers' PNM data and
observed that only $\approx 6.1\%$ of the customers have a ``part of
primary'' ticket and the rest $\approx 93.9\%$ of the customers did
not report anything.

To reduce noise in tickets, we select a subset of customer tickets
that are likely to be caused by network problems. We select the
tickets based on both a ticket's action field and the ticket's
description field.  From the action field, we select tickets that lead
to a ``Dispatch'' action. We assume that the tickets that caused an
ISP to dispatch a technician are likely to be triggered by
network-related problems. From the description field, we select
tickets whose ticket description keywords suggest networking
problems. Examples of such keywords include ``Data Down'', ``Noisy
Line'' and ``Slow Speed''. In the rest of this paper, we refer to
those selected tickets as ``network-related tickets''.

Figure~\ref{f:snr_rate} compares how the ticketing rate of
network-related tickets and all tickets vary with SNR values.  As can
be seen, network-related tickets have higher ticketing rates when SNR
is low, suggesting that the occurrences of those tickets are better
indicators of network faults.

We note that according to the ISP who provided us the datasets,
network-related tickets may also contain non-networking tickets due to
human errors.  A human operator who fills a ticket action or
description field may make a mistake. And a technician may be
dispatched when there is no network fault due to an erroneous
diagnosis.

\PENDING{
Once we can detect failures with PNM data, we also hope to distinguish types of faults from the data.
To gain insights into how to separate a service issue from a
maintenance issue, we manually examined several anomaly patterns by
plotting PNM metrics. % described in \S~\ref{sec:dataset}.
Figure~\ref{f:tx_power_examples} shows an example. In this figure, we
sampled the transmission power levels of devices with three anomaly
patterns from an \FN. The orange \orangedots show the
transmission power levels of three devices that exhibit similar
anomalous patterns in the changes of their transmission powers. When a
noise leaks inside a cable transmission channel, a device increases
its data transmission power to overwhelm the noise. So a sudden
increase in transmission power is an indicator of noise invasion. The
green \greendots show the transmission power levels of two devices that are
not impacted by the noise. The red \reddots show the transmission power
levels of a device that exhibits a different anomalous pattern.}

\PENDING{Figure~\ref{f:tx_power_examples_map} shows the geographic distribution
of the devices in the \FN. We use the same color coding schemes to
plot the devices. The orange \orangedots plot the scrambled
geographic locations of the devices that exhibit similar anomalous
patterns as depicted in Figure~\ref{f:tx_power_examples}. Each red
\reddot shows the scrambled geographic location of a device that
exhibits a distinct anomalous pattern. And the green \greendots show
the locations of the devices that do not exhibit any anomalous
pattern.}

\PENDING{From this data visualization step, we gained the conceptual
understanding that we could use clustering to distinguish a
maintenance issue from a distinct service issue.  In addition, we
  observe that fault detection is independent of clustering, as both
  the healthy devices (the green group) and the unhealthy ones (the
  orange group) form distinct clusters. }

\noindent \textbf{Challenges:} A key question we need to answer is how
to use customer tickets as hints for detecting network
faults. Ideally, if a customer calls only when a network fault occurs,
we could label the PNM records collected around the ticket creation
time as abnormal, and apply supervised learning to learn the PNM
thresholds that suggest a network fault. We have tried several such
machine learning algorithms when we started this project, but found
that this approach did not work well with our datasets. First,
customer calls are unreliable fault indicators. A customer may or may
not call when there is a fault and vice versa. Second, PNM data
contain noise. During a faulty period, some PNM metrics may
occasionally show normal values due to the added noise. Similarly,
even when there is no fault, some PNM metrics may show instantaneous
abnormal values.  Thus, if we use the tickets to label PNM data, it
may introduce many false positives as well as many false negatives. We
found it challenging to tune a machine learning algorithm with this
labeling method. It is even harder to explain the results when we
change a parameter.  Next, we describe how we design \name to use a
simple and customized classifier to address these challenges.

\section{Design}\label{sec:design}

In this section, we describe the design of \name. We first describe
how we reduce the noise in customer tickets and the noise in PNM
data. We then describe a customized classifier that aims to robustly
classify PNM values as normal and abnormal despite the presence of
noise. \PENDING{Next, we introduce how we cluster anomaly patterns to determine network fault types based on the classification results.}  Finally, we describe how an ISP can use 
\sys
to detect network faults and to help diagnose a customer's
trouble call.

\subsection{Reducing Noise in PNM Data}
\label{sec:reduce_noise_time_series_feature}

PNM data measure the instantaneous state of cable's RF channels and
contain noise. An added noise may make a PNM metric take an abnormally
low or high instantaneous value. To address this problem, we treat PNM
data as time-series data and apply statistical models to smooth the
noise and generate additional features for fault detection.

Table~\ref{tab:features_equation} summarizes all the statistical
models we use to process PNM data. For each PNM metric collected at
timestamp $i$ with value $V_i$, we calculate its average, its weighted
moving average (WMA), exponentially weighted moving average (EWMA),
the difference between the current value and its WMA (WMA Diff), and
its variance. We note that the average, WMA, WMA Diff, and variance
values all require a window size as a hyper-parameter. Because we do
not have any prior knowledge on how to set this parameter, we try a
series of window sizes, ranging from \emph{1 day} to \emph{7 days},
incrementing by 1 day at each step.  For the $\lambda$ parameter
required by EWMA, we vary the value of $\lambda$ from \emph{0.1} to
\emph{0.9}, incrementing by 0.1 at each step.  For each PNM metric, we
generate 37 time-series features. We apply this approach to all nine
PNM metrics and totally generate 333 time-series features.  We refer
to them as time-series features.

\begin{table}[!tb]
\newcommand{\tabincell}[2]{\begin{tabular}{@{}#1@{}}#2\end{tabular}}
\centering
\begin{tabular}{c|c}
% \toprule
\textbf{Model}    & \textbf{Equation}  \\ \toprule

Average   & $AVG_i=\frac{V_i+V_{i-1}+\dots+V_{i-win+1}}{win}$    \\ \hline

WMA       & $WMA_i=\frac{win \cdot V_i+ (win-1) \cdot V_{i-1}+\dots+1\cdot V_{i-win+1}}{win\cdot (win-1)/2}$     \\ \hline

EWMA      & \tabincell{c}{$EWMA_1 = V_1$ \\ $EWMA_i = \lambda \cdot V_i + (1-\lambda)EWMA_{i-1}$}  \\ \hline

WMA Diff  &  $V_i - WMA_i$\\ \hline

Variance & $VAR_i = \frac{1}{win}\sum_{k=i-win+1}^i(V_k-AVG_i)^2$\\
% \bottomrule
\end{tabular}
\caption{\small\textbf{This table summarizes the statistical models
    we use to generate the time-series features. (WMA: Weighted Moving
    Average, EWMA: Exponentially Weighted Moving
    Average.)\label{tab:features_equation}}}
\end{table}

\subsection{Determining A Fault Detection Threshold}

After we reduce noise in both the customer tickets and the PNM data,
we aim to determine a threshold for each PNM metric that indicates
network faults. We note that there is no explicit definition of what a
network fault is. Instead, we choose to use the network conditions
that are likely to cause a trouble call to approximate a network
fault. With this approximation, we may not detect minor issues that do
not warrant a trouble call. We argue that this design is advantageous,
because it allows an ISP to prioritize its resources to fix the
customer-impacting problems.

In the case of SNR, if we choose too high a value as a fault detection
threshold, an ISP may become too proactive, fixing minor problems that
many customers may not care, which we refer to as false positives. If
we choose too low a value, an ISP may miss opportunities to
proactively repair a problem before a customer calls, which we refer
to as false negatives.

We aim to design an algorithm that minimizes both false positives and
false negatives.  From our investigation in
\S~\ref{sec:design_rationale}, we see that different values of a PNM
metric have different likelihood to concur with a trouble
ticket. Inspired by this observation, we use the ticketing rate as a
metric to help choose a fault detection threshold.  Our intuition is
that the customer ticketing rate during a faulty period should be
higher than a normal period when there is no fault. Therefore, for
each feature $f$ generated from a PNM metric, we determine a threshold
value $thr_f$ such that $thr_f$ maximizes the ratio between the
ticketing rate in the time periods when a network fault exists and the
time periods when there is no fault. We refer to this ratio as the
ticketing rate ratio.

Specifically, we search through the range of values of a feature $f$
in small steps. At each step $s$, we consider the value of the feature
$f_s \in [f_{min}, f_{max}]$, as a candidate for the threshold.  We
then compare the value of $f$ at a PNM data collection point with
$f_s$, and label the collection time period as abnormal or normal,
based on whether the value of $f$ is below or above the candidate
threshold value $f_s$. For some features such as the average SNR,
below the threshold is abnormal. For other features, the opposite is
true.  After determining each collection period as normal or abnormal,
we count the number of network-related tickets occurred in the normal
and abnormal periods respectively and divide them by the normal and
abnormal time periods determined by $f_s$. We then compute the
ticketing rate ratio: $TRR(f_s)$. The threshold value $thr_f$ is
chosen as the value of $f_s$ that maximizes the ticketing rate ratio
$TRR(f_s)$.

We also note that for features following a normal distribution such as
Rx Power, we choose to use two threshold values to determine whether a
collection period is normal or abnormal. 

\begin{figure}[tb!]
\centering
    \includegraphics[width=.5\textwidth]{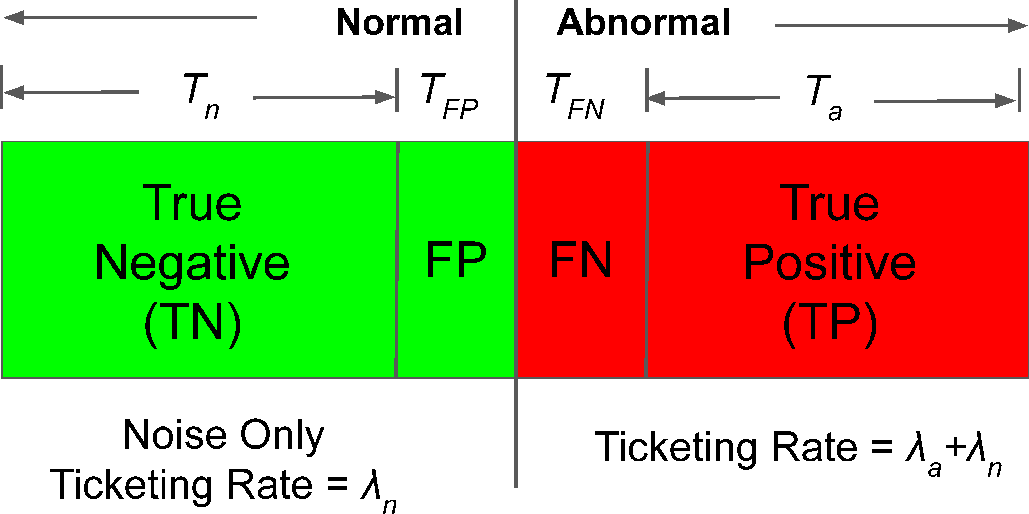}
    \caption{\small\textbf{Analysis of ticketing rate ratio.}}
    \label{f:analysis}
\end{figure}

We now explain why choosing a threshold that maximizes the ticketing
rate ratio may help minimize the false positives and false
negatives. The entire timeline can be divided into two subspaces: the
normal (no fault) and the abnormal (with fault) periods.  Ideally, the
normal sub-space should not receive any trouble ticket. In practice,
there is always a ticketing noise. We assume a uniformly distributed
ticketing noise with the rate $\lambda_n$ spreads the whole
space. Similarly, we assume an additional uniformly distributed
ticketing rate that occurs only in the abnormal sub-space and denote
it as $\lambda_a$.

A threshold value $thr_f$ of a feature also divides the timeline into
two subspaces: normal and abnormal. The first subspace includes a true
negative part $T_n$ and a false negative part $T_{FN}$, where an
abnormal period is erroneously considered as normal.  The second
subspace includes a true positive part $T_a$ and a false positive part
$T_{FP}$, where a normal period is considered abnormal.  The ticketing
rate ratio determined by the threshold $thr_f$ can be computed as
follows:

$$D(T_n, T_{FP}, T_{FN}, T_a) =
\frac{\frac{\lambda_nT_{FP}+(\lambda_a+\lambda_n)T_a}{T_a+T_{FP}}}{\frac{\lambda_nT_n+(\lambda_a+\lambda_n)T_{FN}}{T_n+T_{FN}}}$$

Both the numerator and denominator can be regarded as a weighted
average of $\lambda_n$ and $\lambda_a+\lambda_n$, with the time period
lengths as the weights. Because $\lambda_a+\lambda_n>\lambda_n$ always
holds, we can show that the derivatives of $D$ over the false
positives $T_{FP}$ and the false negatives $T_{FN}$ are non-increasing:

$$\frac{\partial D}{\partial T_{FP}} < 0~~{\rm and}~~\frac{\partial D}{\partial T_{FN}} < 0$$

\noindent Therefore, because $T_{FP}$ and $T_{FN}$ are non-negative,
the ticket rate ratio is maximized when both false positives and false
negatives are zero:

$$D_{max} = \lim\limits_{\substack{T_{FP}\rightarrow0\\T_{FN}\rightarrow0}}D = \frac{\lambda_a}{\lambda_n}+1$$

\subsection{Feature Selection}\label{s:feature_selection}

We have a total of more than three hundred time-series features and it
is unlikely they are all useful indicators of network faults. To find
the relevant features, we only select the features with high ticketing
rate ratios from each PNM metric. Specifically, among the same type of
features derived from a PNM metric with different hyperparameters, we
choose the one with the highest ticketing rate ratio as the
representative feature.  For each representative feature derived from
the same PNM value, we choose the top two with the highest ticketing
rate ratios.  Finally, among the remaining candidates, we choose the
top $N$ features that have the highest ticketing rate ratios. We
determine the number of features $N$ based on the desired ticketing
rate ratios, \precision, and \recall as we soon describe in
\S~\ref{sec:eval_metric}.

Table~\ref{tab:features_rate} shows the top five features we used and
their ticketing rate ratios calculated from our training sets
(Section~\ref{sec:evalSetup}). The name of each feature consists of
the raw PNM metric, the statistical model we apply to the metric, and
the parameter. For example, the \emph{snr-var-2} means the variance of
SNR with a \emph{2-day} window size. We note that all features have a
high ticketing rate ratio and we expect them to effectively detect
network faults.

\begin{table}[!tb]
\centering
\begin{tabular}{c|c}
% \toprule
\textbf{Features}              & \textbf{Ticketing Rate Ratio} \\
\toprule
snr-var-2             & 14.49\\ \hline
uncorrected-var-1     & 7.66 \\ \hline
rxpower-wma-diff-4    & 5.31 \\ \hline
t3timeouts-wma-diff-1 & 4.93 \\ \hline
t4timeouts-var-1      & 4.18 \\
% \bottomrule
\end{tabular}
\caption{\small\textbf{Top 5 features and their ticketing rate ratio. \label{tab:features_rate}}}
\end{table}

\subsection{Combining Different Features}
\label{sec:or_combining}

Different PNM features may detect different types of network
faults. Therefore, we build the final classifier by combining the
detection results of all selected features. As long as one selected
feature considers a PNM collection period abnormal, we classify the
collection period as abnormal. For each selected feature, we have
already chosen a threshold that maximizes the ticketing rate
ratio. Therefore, we expect that combining the results of all selected
features will also provide a high ticketing rate ratio.  We evaluate
the results of our classifier in \S~\ref{sec:eval_metric}.

\subsection{\PENDING{Anomaly Clustering}}
\label{sec:design:telapart}

After identifying features for fault detection, we design a clustering algorithm for distinguishing maintenance and service issues by grouping detected anomaly events.
  
The features used for grouping anomaly events are different from the ones used for detection. \PENDING{When detecting anomalies, \sys relies on time-series features to reduce noises (\S\ref{sec:reduce_noise_time_series_feature}). Essentially, the time-series features can make \sys's detection more noise robust because it involves more data points. As for anomaly clustering, a cluster naturally contains multiple devices and thus, for each timestamp, there are already multiple data points within a cluster. As a result, PNM features can be used directly for anomaly clustering. In terms of the selected PNM features,} some detection features contain
the instantaneous values measured at the data collection times, while
others are cumulative values (e.g. codeword error counters) over
time. We find that the instantaneous metrics, including SNR, Tx power,
and Rx power, are effective features for grouping devices with shared
maintenance issues together. 

On the other hand,
cumulative metrics, although effective in detecting
anomalies, are not as effective as clustering
features. % for various reasons.  
For example, the values of codeword
error counters are affected by whether users actively use the Internet
or not. Devices that share the same maintenance issue may
or may not have highly correlated codeword error counters if the
subscribers' usage patterns differ. Hence, \sys % this work
uses only
instantaneous metrics for clustering: SNR, Tx power, and Rx power. Among them, Tx and
Rx powers are statistically correlated. Finally, we retain two independent
features: SNR and Tx power for clustering. 

We employ the average-linkage hierarchical
clustering algorithm~\cite{yu2015hierarchical} to group anomaly events.  At a high level, this clustering
algorithm works as follows. For each feature $f$ we selected, the
clustering algorithm aims to group devices with similar feature
vectors together until the similarity between groups of devices falls
below a threshold $s_f$. Specifically, it first treats each device
(described by a feature vector) as a single cluster. Second, it
calculates the similarity between every pair of clusters and finds two
clusters with the highest similarity value. The similarity between two
clusters is calculated by averaging all similarity values between
pairs of devices in the two clusters. Third, the algorithm merges the
two clusters with the highest similarity value into a single
cluster. Next, the algorithm repeats the second and third steps until
only one cluster is left or the highest similarity value between any
two clusters is less than the similarity threshold $s_{f}$. Finally, the
algorithm outputs the clusters that have not been merged.

The similarity threshold $s_f$ for each feature $f$ is an important
hyper-parameter and \sys's performance is sensitive to its value.  If
we set the threshold too high, \sys may separate devices that are
affected by the same network fault into multiple clusters. Conversely,
if we set the threshold too low, it may group devices that are
affected by different maintenance issues into the same cluster.

How do we choose a proper similarity threshold? If we had labeled
training data, we could use the grid-search
method~\cite{lavalle2004relationship} to iterate over possible values and set the threshold that
minimizes clustering errors. 
Lacking  labeled data, we instead use customer ticket statistics to
guide the search for the similarity threshold. Our insight is that if
\sys correctly identifies groups of devices that are impacted by the
same maintenance issue, then on average, we should observe a higher
fraction of maintenance tickets reported by these groups of devices
than other devices. In contrast, if \sys partitions the cable devices
rather randomly, then we should not observe significant statistical
differences of the reported maintenance tickets among different
groups.

Motivated by this insight, we naturally evolve the \emph{ticketing rate} to \emph{maintenance ticketing rate} and devise the following mechanism to set
the similarity threshold $s_f$ for each PNM feature we use. We
partition the PNM dataset we have into a training set and a testing
set. For each data point $i$ in the training set and for each possible
value of $s_f$, we use \sys to diagnose whether a device $j$ is
impaired by an infrastructure fault and the type of fault. 
If \sys considers a device experiencing a
maintenance issue, we mark this collection period of this device as a
maintenance event. We use $I_{i,j}$ to denote the length of the data
collection interval between data points $i$ and $i-1$ of device $j$. Similarly, if
\sys considers a device experiencing a service issue, we mark the
collection period of the device as a service event. We then count the
number of maintenance tickets reported by all devices
during all collection periods that are marked as maintenance issues
and compute a maintenance ticketing rate during maintenance events as
\begin{equation}
  R_{m,M} = \frac{K_{m,M}} {\sum_{i,j} I^{M}_{i,j}}
  \label{eq:rt_m}
\end{equation}
where $K$ denotes the number of tickets, $R$ denotes the ticketing
rate, the first subscript $m$ denotes maintenance tickets, and the
second subscript $M$ denotes a diagnosed maintenance issue, and
  $I^{M}_{i,j}$ is the length of a collection period that is marked as
 experiencing a maintenance issue.

We also count the number of maintenance tickets $K_{m,S}$ reported by
all devices during all collection periods that are marked as service
issues.  We compute a maintenance ticketing rate during service events
as
\begin{equation}
  R_{m,S} = \frac{K_{m,S}}{\sum_{i,j} I^{S}_{i,j}}
  \label{eq:rt_s}
\end{equation}
where $S$ indicates a diagnosed service issue, and
$I^{S}_{i,j}$ is the length of a collection period marked as
experiencing a service issue.  We define the maintenance Ticketing
Rate Ratio ($TRR_m$) as
\begin{equation}
  TRR_m = \frac{R_{m,M}}{R_{m,S}}
  \label{eq:trrm}
\end{equation}
For each feature $f$ \sys uses, we use grid-search to find the
similarity threshold value $s_f$ that maximizes $TRR_m$.
% In Appendix~\ref{sec:cluster_proof}, we prove 
It can be proven\footnote{We skip the formal proof because of the limit of space.} that the $s_f$ maximizing $TRR_m$
yields the
optimal clustering result: it minimizes both false positives (\ie, a device without any maintenance issues is detected
as with a maintenance issue) and false negatives (\ie, a device with a maintenance issue is
detected as without any maintenance issues). Intuitively, based on \anonisp's fault diagnosis
process, maintenance tickets contain fewer false positives than
service tickets. Thus, if we assume the operator-labeled
maintenance tickets approximate the unknown but existing ground truth
of maintenance events and \sys's fault detection mechanism is
accurate, then the maintenance ticketing rate during maintenance
events approximates \sys's true positives and the maintenance ticketing
rate during service events approximates \sys's false negatives. Maximizing
the ratio of the two ticketing rates leads to high true positives
and  low false negatives.

\subsection{ISP Deployment}
\label{sec:design_isp}

An ISP can use \name in two ways: proactive network maintenance for
predicted trouble tickets and diagnosing the root cause of a trouble
ticket when receiving a call. In this section, we describe the
algorithms for an ISP to decide when to send out a repair technician
proactively and how to diagnose the root cause.

\begin{figure}[tb!]
\centering
    \includegraphics[width=.5\textwidth]{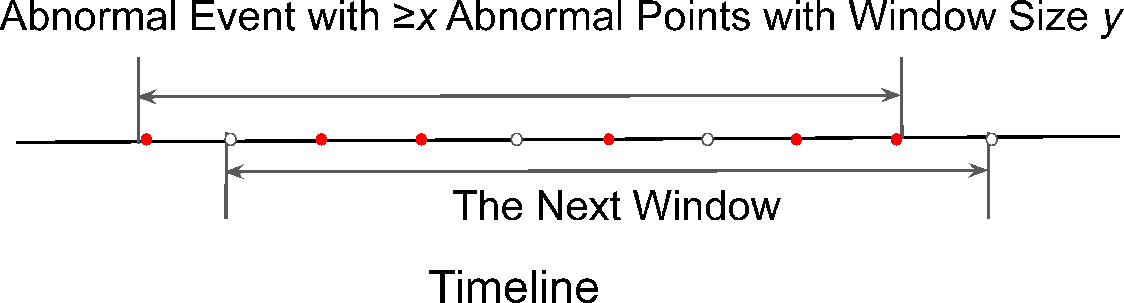}
    \caption{\small\textbf{This figure explains the sliding window
        algorithm. When the number of abnormal points within a sliding
        window exceeds a threshold, the window is considered to be
        abnormal. An abnormal event is given by merging the abnormal
        windows.}}
    \label{f:sliding_window}
\end{figure}

\name's classifier can monitor an ISP's network continuously. It can
output a normal and abnormal decision when a PNM record is collected
from a customer's modem. However, due to the existence of noise and the
intermittent nature of some faults, if an ISP makes a dispatch decision
whenever it observes an abnormal PNM data point, it may lead to many
false positives. To address this problem, we design a sliding window
algorithm for an ISP to make a dispatch decision. The high-level idea
of this algorithm is that an ISP should only dispatch a technician
after a fault persists.

Figure~\ref{f:sliding_window} explains this algorithm. The algorithm
takes two parameters: $y$ and $x$, where $y$ is the size of the
window, and $x$ is the number of abnormal data points detected in the
window. When an ISP collects a new PNM record, it looks back to a
window size $y$ of collection points. If $x$ out of $y$ data points are
considered as abnormal, then the ISP should dispatch a technician to
examine and repair the network. \PENDING{The type of the dispatched technician can be determined by the fault type detected by \sys's clustering algorithm.}

An ISP can determine the parameters $x$ and $y$ based on the false
positives and false negatives it is willing to tolerate. The ISP can
estimate the values of false positives and false negatives from its
historic PNM data and ticket data. Therefore, choosing those
parameters only requires an ISP to train \name using its own PNM and
ticket data and does not require tuning.  In \S~\ref{sec:eval_metric},
we use our datasets to show how an ISP can effectively choose the
parameters $x$ and $y$.

Similarly, an ISP can use \name to help diagnose the root cause of a
call. When it receives a trouble call, if the customer complains about
a performance problem, and the ISP sees that in the past collection
window of size $y$, there exists $x$ abnormal collection points, the
ISP can conclude that the trouble is likely to be caused by a network
problem\PENDING{, furthermore, whether it is a service or maintenance issue}.

\section{Evaluation}
\label{sec:eval}

In this section, we describe how we evaluate \name's
performance.

\subsection{Establishing Evaluation Metrics}
\label{sec:eval_metric}

Ideally, we would like to deploy \name on a real cable ISP and measure
how it reduces the number of trouble tickets over a long term.  It is
our future work to conduct such a real-world experiment. In this work,
we aim to estimate how many trouble tickets \name would reduce were it
deployed on our collaborating ISP.

To do so, we emulate the sliding window algorithm described in
\S~\ref{sec:design_isp} using our test dataset. We start from the
beginning of the test dataset. If there are $x$ abnormal points
detected in a window size of $y$, we mark it as the beginning of a
fault. We then move the window to the next data point. When the number
of abnormal points falls below $x$, we mark it as the end of a
fault. If there is a trouble ticket occurred during a fault, we
consider this fault detection as a true fault. We note that if we
detect a fault simultaneously within multiple customers, as long as
one customer reports a ticket, we consider it a true fault.  We assume
that if an ISP dispatched a repair technician when it detected the
onset of the fault, it could have avoided the trouble ticket.  We
define \emph{\precision} as the number of true faults divided by the
total number of detected faults. We define \emph{\recall} as the
number of trouble tickets occurred during a detected fault divided by
the total number of network-related trouble tickets.

It is not sufficient to use only \precision and \recall to gauge
\name's performance. This is because if \name detects the entire time
period that spans the test dataset as a faulty period, it will achieve
100\% \recall and \precision. To avoid this pitfall, we also use the
normalized ticketing rate, which is defined as the ticketing rate in
all faculty periods normalized by the ticketing rate of the time
period that spans the test dataset.  If \name erroneously detects the
entire time period as abnormal, it will achieve a low normalized
ticketing rate close to 1.

\begin{figure}[tb!]
\centering
    \includegraphics[width=\columnwidth]{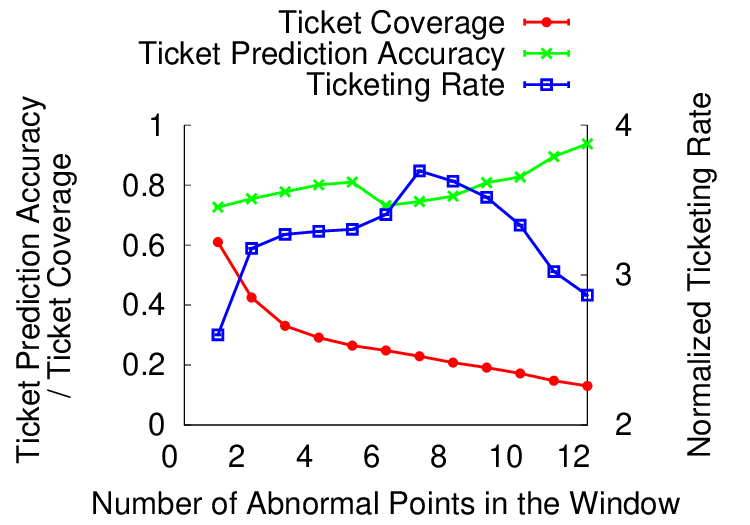}
    \caption{\small\textbf{This figure shows the ticket detection
        accuracy, the ticket coverage, and the normalized ticketing
        rate of the sliding window algorithm with different
        parameters.}}
    \label{fig:slide}
\end{figure}

\noindent \textbf{How an ISP chooses the sliding window parameters:}
In practice, an ISP can use a training set to determine the threshold
values of \name's classifier. It can use the \precision, \recall, and
the normalized ticketing rate obtained from a validation set to choose
the combination of the sliding window parameters.

We show an example in Figure~\ref{fig:slide}. In this example, we
choose a window size of 12 data points ($y$ = 12), which is roughly
two days long. We then measure the \precision, \recall, and the
normalized ticketing rate when the number of abnormal points $x$
varies from $0$ to $12$. As can be seen, when $x$ is around 8, the
sliding window algorithm achieves a high normalized ticketing rate, a
relatively high \precision $80\%$, and a \recall around $20\%$. Since
avoiding false dispatches is more important than predicting all
trouble tickets, an ISP can choose (8, 12) as its sliding window
parameters for fault detection.

\begin{figure}[tb!]
\centering
    \includegraphics[width=\columnwidth]{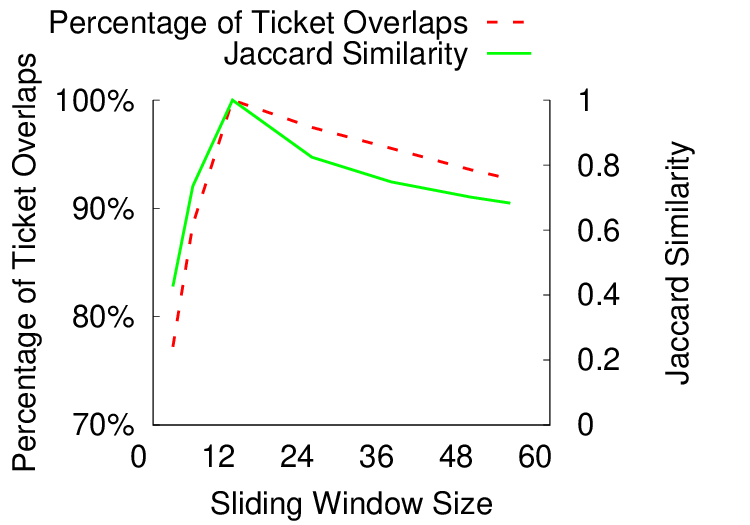}
    \caption{\small\textbf{This figure shows what percentage of
        tickets detected by different window sizes overlap with those
        detected by a window size of 12 and the Jaccard similarity
        between the faulty periods detected by different window sizes
        and those detected by a window size of 12.}}
    \label{fig:window_overlap}
\end{figure}

We have tried different sizes of the sliding window, ranging from one
to 60 data points. For each window size, we use the above method to
choose the parameter $x$ such that both the \precision and the
normalized ticketing rate are high, and the \recall is above a minimum
threshold $15\%$. We compare the tickets and the faulty periods
detected by different window parameters. We use the Jaccard similarity
metric~\cite{levandowsky1971distance} to measure the overlaps of
faulty periods detected by different window parameters. As can be seen
in Figure~\ref{fig:window_overlap}, 90\% of the tickets detected by
windows larger than 12 overlap; and the faulty periods detected by
them have a Jaccard similarity larger than 60\%. This result suggests
that different window parameters are likely to detect the same sets of
faults, and the performance of \name is not sensitive to the window
parameters.

\subsection{Experiment Setup}
\label{sec:evalSetup}

After we establish the evaluation metric, we train and evaluate \name
on a 50-machine Linux cluster with $40\sim 512$ GB RAM and $8\sim 48$
core running Ubuntu 18.04. \Name is trained on five-month data from
01/06/2019 to 05/30/2019 and tested with three-month data from
06/01/2019 to 08/31/2019. 

\label{sec:bxe}

\paragraph{Comparing with the Existing Work:} We compare \name's
performance with two existing methods. One is from our collaborating
ISP, AnonISP, which uses a visualization tool that colors different
ranges of PNM values for an operator to manually monitor its networks'
conditions.  AnonISP's tool has two manually configured thresholds for
several raw PNM values and therefore has three fault indication
levels: normal (green), marginal (yellow), and abnormal (red). We
compare AnonISP's tool against \name with these thresholds and regard
both yellow and red levels as network faults, as the ISP's experts
usually do.

Another tool from the industry uses Comcast's scoreboard
method~\cite{larry2016comcast}. Comcast is considered as the leading
company in the area of PNM research. They developed a method that
compares a PNM metric to a threshold and assigns a score to each
comparison result. If the sum of the comparison scores  exceeds a
threshold value, then the method considers there is a fault in the
network. 

Since both AnonISP and Comcast's tool detect a fault using a single
PNM data record, we apply the sliding window algorithm to both tools
for a fair comparison.

\paragraph{Comparing with Machine Learning Techniques:}
We also compare the performance of \name with three classical machine
learning algorithms: Decision Tree (DT)~\cite{quinlan2014c4}, Random
Forest (RF)~\cite{breiman2001random} and Support Vector Machine
(SVM)~\cite{hearst1998support}.  Since these algorithms require
labeled data, we label the PNM data with tickets. Each ticket has a
creation time and a closed time. We label the PNM data collected
between this time interval as positive samples and other data as
negative samples. We generate 47,518 positive samples and the same
number of abnormal samples as our training set to train the machine
learning models and evaluate them with the same evaluation metrics.

Table~\ref{tab:ml_comparing} shows the \precision, the \recall, and the
normalized ticketing rate of different methods. As can be seen, \sys
achieves the highest ticket prediction accuracy and the highest
normalized ticketing rate among all methods. Its ticket coverage is
lower than that of AnonISP. However, this is because AnonISP detects
too many false faults, as shown by its low ticket prediction accuracy.
We note that all three machine learning algorithms require a long
training time, as each has multiple parameters to tune. The results we
present here are the best ones after many rounds of tuning. When we
started this project, we started with those algorithms, but abandoned
them due to the challenges to tune them and to explain the results
when certain parameters are changed.

\begin{table}[H]
\centering
\begin{tabular}{c|c|c|c}
\multirow{3}{*}{Methods} & Ticket & Ticket & Normalized \\
& Prediction & Coverage & Ticketing Rate\\
& Accuracy & &\\
\toprule
\name  & 81.92\% & 22.99\% & 3.55 \\ \hline
Decision Tree     & 68.93\% & 15.53\% & 2.52 \\ \hline
SVM    & 75.64\% & 12.54\% & 2.02 \\ \hline
Random Forest     & 73.14\% & 14.21\% & 2.24 \\ \hline
Comcast & 23.48\% & 2.21\% & 1.18 \\ \hline
AnonISP's tool & 10.04\% & 25.13\% & 0.98 \\ 
\end{tabular}
\caption{\small\textbf{Performance of different methods \label{tab:ml_comparing}}}
\end{table}

\subsection{Detected Tickets Statistics}
\begin{figure}[tb!]
\centering
    \includegraphics[width=\columnwidth]{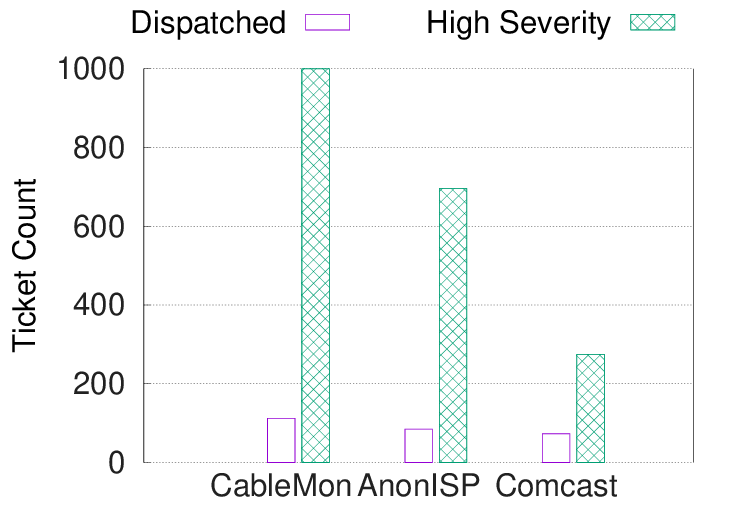}
    \caption{\small\textbf{This figure shows the number of different
        types of tickets detected by different methods.}}
    \label{f:tk_stats}
\end{figure}

\begin{figure}[t!]
\centering
\subfigure[CDF]{
    \includegraphics[width=\linewidth]{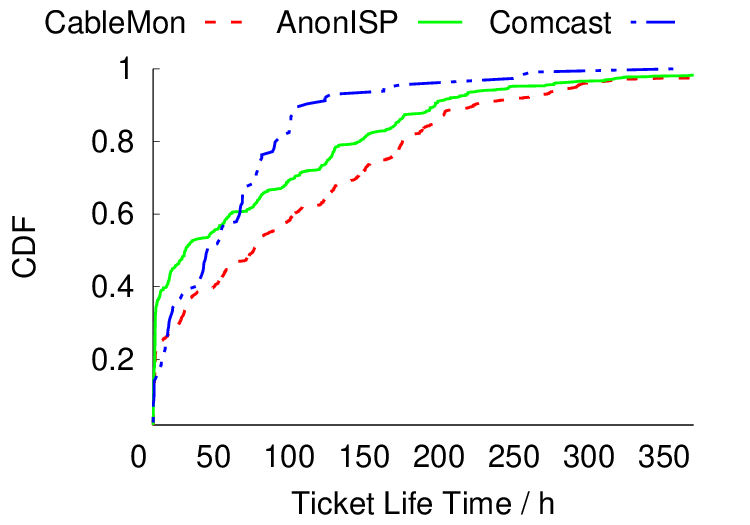}
    \label{f:tk_epoch_cdf}
} \hspace{-15pt}
\subfigure[Mean and Median]{
    \includegraphics[width=\linewidth]{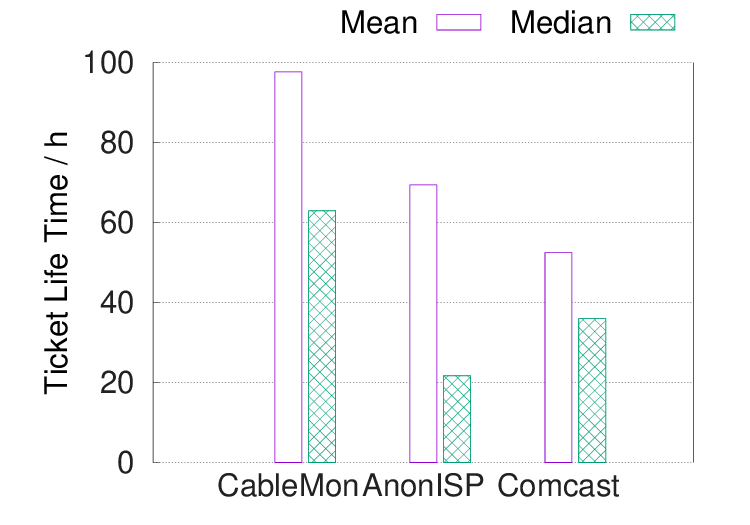}
    \label{f:tk_epoch}
}
\caption{\small\textbf{The figures show the CDF, mean, and median of
    the life time of tickets predicted by different methods. A longer
    ticket life time indicates that the problem that triggered the
    ticket takes a longer time to fix.}}
\end{figure}

\begin{figure}[t!]
\centering
\subfigure[CDF]{
    \includegraphics[width=\linewidth]{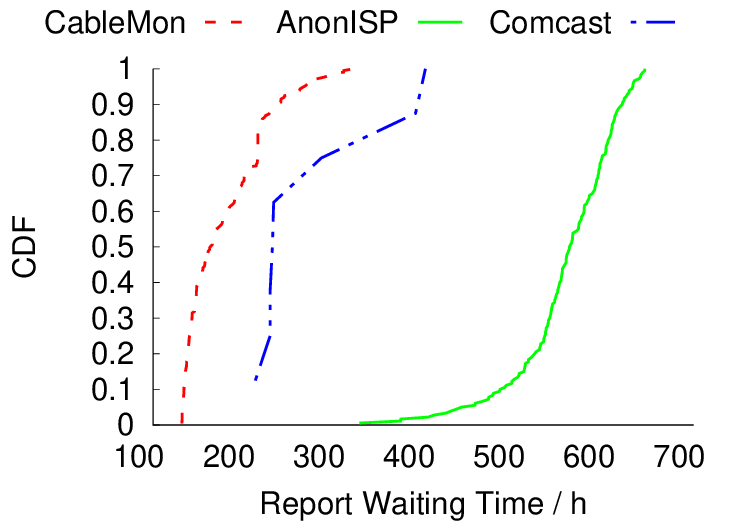}
    \label{f:tk_report_cdf}
} \hspace{-15pt}
\subfigure[Mean and Median]{
    \includegraphics[width=\linewidth]{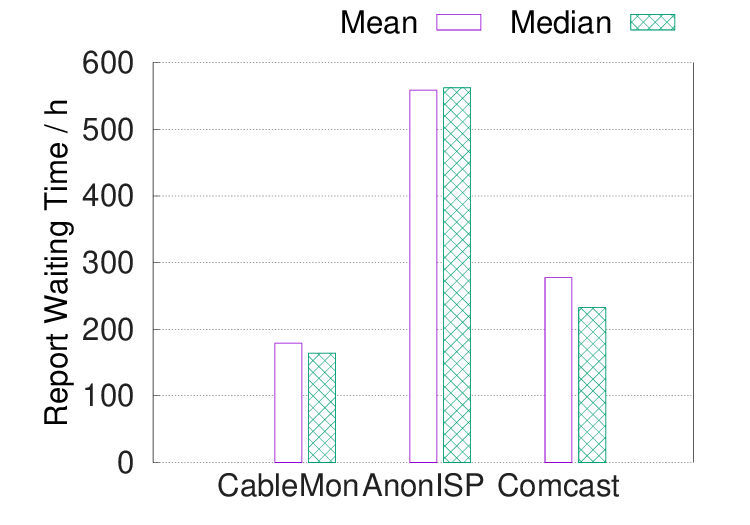}
    \label{f:tk_report}
}
\caption{\small\textbf{The figures show the CDF, mean, and median of
    the report waiting time of tickets predicted by different
    methods. A shorter report waiting time indicates that the problem
    triggered by the ticket is more urgent.}}
\end{figure}

\begin{figure}[tb!]
\centering
    \includegraphics[width=\columnwidth]{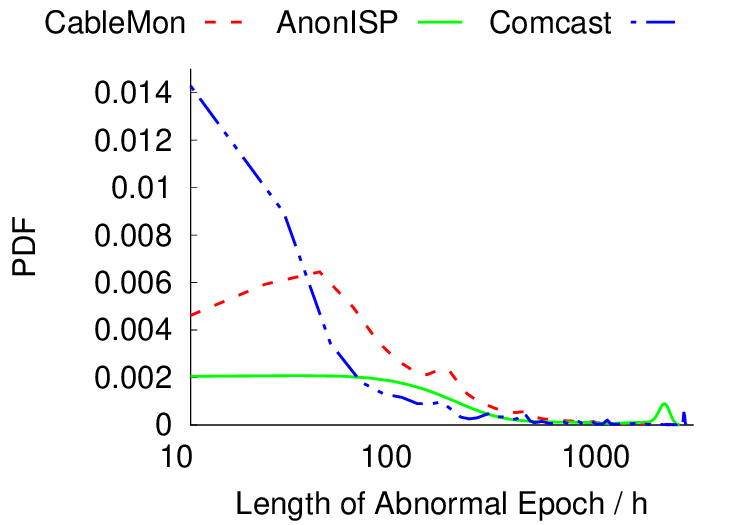}
    \caption{\small\textbf{This figure shows the PDF of the
        length of a detected fault.}}
    \label{f:period_epoch}
\end{figure}

To further analyze the detected tickets, we examine the tickets
detected by \sys and existing ISP tools according to the ticket action
and description fields. We omit the results of the machine learning
algorithms for clarity. The characteristics of the tickets detected by
those algorithms are similar to those of \sys, but they have lower
ticket detection accuracy and coverage.  Figure~\ref{f:tk_stats} shows
the number of different types of tickets detected by different
methods. As can be seen, \name can detect a significantly greater number
of dispatched and higher severity tickets than the two existing ISP
tools.

Figure~\ref{f:tk_epoch_cdf} shows the distribution of a detected
ticket's life time, and figure~\ref{f:tk_epoch} shows the average and median life
time of a detected ticket. A ticket's life time is defined as the time
between a ticket is created to the time a ticket is closed.  As can be
seen, the tickets detected by \name have longer life times,
suggesting that \name detects the problems that take longer to
resolve.

We also measure the time elapsed from when a fault is detected to when
a ticket is created. We refer to this time as ``Report Waiting Time.''
Figure~\ref{f:tk_report_cdf} shows the cumulative distribution of the
report waiting time of different methods, and Figure~\ref{f:tk_report}
shows the average and median report waiting time of different methods. As can be
seen, \Name's report waiting time is also significantly shorter than
that of other methods, indicating that its detected faults lead to
customer trouble tickets faster than those detected by other methods.

Finally, we measure the distribution of a fault detected by different
methods. Figure~\ref{f:period_epoch} shows the PDF of the length of a
fault detected by different methods.  As can be seen, \name detected
faults tend to last a moderate period of time. The highest probability
density is slightly less than 100 hours (roughly four
days). Comcast's tool detects many faults that last less than one day,
shorter than what a typical network repair action takes. This result
suggests that many of the detected faults could be false positives.
The faults detected by AnonISP's tool have a wide range of life span,
from very short faults to very long faults ($>$500 hours), which are
outside the normal range of repair actions.  Again, this result
suggests that many of the detected faults could be false positives.

\subsection{\PENDING{Distinguish Maintenance and Service Issues}}

We started the inspection by choosing  50 maintenance
tickets and used
  the tickets' start and close time to guide the search for anomalous
  patterns. We were able to obtain 16 groups of maintenance issues that
  impact nearly 700 devices. Since we must inspect all devices sharing
  the same fiber optical node for each maintenance ticket, we were also able to identify
  113 devices that were affected by service issues. We carefully
  verified the labeling results with \anonisp's experts to guarantee
  the labeling accuracy.
  
This manual labeling process is cumbersome and error-prone.  It
took two-person-week to obtain these labels.  We intentionally did not
expand into more labels to ensure the labeling accuracy. We
  note that this manually labeled set covers only a small fraction of
  anomalous patterns and is not suitable for training a high-quality
  classifier.
  
We run the clustering algorithm on \pnmdata we label and compare their cluster
results with our labeled results.  We choose two widely used metrics for evaluating each clustering algorithm: the Rand
Index (RI)~\cite{rand1971objective} and the Adjusted Rand Index
(ARI)~\cite{hubert1985comparing}.

We compute RI by comparing the partitions produced by a clustering
algorithm with the ground truth partition.  If two devices are in the
same cluster in both partitions, we count it as a true positive
($TP$).  Conversely, if two devices are in the same subset in the
partition produced by a clustering algorithm, but they are in
different subsets in the ground truth partition, we count it as a
false positive ($FP$). True negatives ($TN$) and false negatives
($FN$) are defined accordingly. RI computes the fraction
of true positives and negatives divided by the total pairs of devices:
$\frac{TP+TN}{TP+TN+FP+FN}$. Its maximum value is 1. The higher the
RI, the better the clustering result.  ARI adjusts for the random
chances that a clustering algorithm groups two devices in the same
cluster by deducting the expected RI ($\mathbb{E}(RI)$) of a random
partition: $\frac{RI-\mathbb{E}(RI)}{1-\mathbb{E}(RI)}$.

Within the architecture of \sys, the average linkage hierarchical clustering algorithm is employed to categorize devices affected by the same network anomaly. A salient challenge is the indeterminacy of the distinct pattern count. Given this inherent uncertainty, clustering algorithms that mandate the specification of the number of clusters, represented by the hyper-parameter $k$, are inherently inconsistent with our design objectives. In the context of \sys, it is imperative to employ algorithms capable of discerning the optimal number of clusters autonomously, circumventing the limitations presented by the need for predefined cluster counts. Therefore, we compare \sys's clustering algorithm choice with three popular clustering algorithms that are not contingent on the predefined $k$ value, including DBSCAN, single-linkage, and complete-linkage clustering algorithms.

\begin{table}[t!]
\scriptsize
% \small
\centering
\begin{tabular}{c c c c c}
% \toprule
                    & \makecell[c]{Average \\ Linkage} & DBSCAN & \makecell[c]{Single \\ Linkage}  & \makecell[c]{Complete \\ Linkage} \\ \toprule
% Purity              & 0.96  & 0.88 & 0.88 & 0.94 \\ \midrule
RI                  & 0.91  & 0.84 & 0.83 & 0.83 \\ \midrule
ARI                 & 0.83  & 0.65 & 0.64 & 0.66 \\ % \bottomrule
\end{tabular}
\caption{\small\textbf{Rand Index and Adjusted Rand Index for various clustering algorithms.}\label{tab:manual_label_result}}
\end{table}

Table~\ref{tab:manual_label_result} shows the comparison results. \sys
achieves an RI of 0.91 and an ARI of 0.83, respectively. \sys's choice outperforms other clustering algorithms. 

To further demonstrate \sys's effectiveness of determining fault types, we collaborate with \anonisp to perform a field test. \sys is wrapped as a service deployed with Docker that can be called by \anonisp's field team. We elaborately designed the API of the \sys service such that the service can be incorporated into \anonisp's workflow. 

We received a summary but no details from the ISP, confirming \sys's effectiveness: ``(we) evaluated the performance of the clustering methodology produced by (your) team and found it effective at the task of classifying defects as service or maintenance''\footnote{Sensitive names are hidden for anonymity.}.

\section{Related Works}

Previous work measured the reliability of broadband networks.  The
Federal Communications Commission launched the Measuring Broadband
America (MBA) project~\cite{fcc2011} since 2010.  Bischof
\ea~\cite{bischof2017characterizing} showed that poor reliability will heavily
affect user traffic demand. Padmanabhan \ea~\cite{padmanabhan2019residential} demonstrated that the outages of
broadband networks tend to happen under bad weather
conditions. Baltrunas \ea~\cite{baltrunas2014measuring} also measured
the reliability of mobile broadband networks.

Network fault diagnosis has attracted much attention from the
community for a long time. Many approaches from the industry, especially
the cable industry, set manual thresholds for certain measured metrics
to detect network outages. Amazon~\cite{ferraris2012evaluating} used a
fixed threshold to monitor the condition of its cloud services. Zhuo
\ea~\cite{zhuo2017understanding} treated packet loss as a fault
indicator and showed the correlation between Tx/Rx Power and packet
loss rate.  They again use manually set thresholds to detect network
faults. Lakhina \ea~\cite{lakhina2004diagnosing} proposed the first
framework that applied Principal Component Analysis (PCA) to reduce
the dimension of network traffic metrics. Huang
\ea~\cite{huang2007network} showed that Lakhina’s framework works well
with a limited number of network devices, but has performance issues
on larger datasets. Moreover, Ringberg
\ea~\cite{ringberg2007sensitivity} pointed out that using PCA for
traffic anomaly detection is much more tricky than it appears. Besides
PCA, many other statistical models are applied to network anomaly
detection. Gu \ea~\cite{gu2005detecting} measured the relative entropy
of the metrics and compared them to the
baseline. Subspace~\cite{li2006detection} is introduced to deal with
high-dimensional and noisy network monitoring data. Kai
\ea~\cite{kai2009network} used Expectation–Maximization (EM) algorithm
to estimate the parameters of their model and obtain the upper or
lower bound of the common metric values. Independent Component
Analysis~\cite{palmieri2014distributed}, Markov Modulated
Process~\cite{paschalidis2009spatio}, and Recurrence Quantification
Analysis~\cite{palmieri2010network} are also introduced to find the
anomaly points in time series data.  These methods aim to detect
sudden changes in data. Differently, \sys uses customer tickets as
hints to label the input data and uses the ticketing rate ratio to
select relevant features.

Recently, machine learning has been used for network anomaly
detection. Leung \ea~\cite{leung2005unsupervised} designed a
networking anomaly detection system using a density-based clustering
algorithm, which obtained an accuracy as 97.3\%. Dean
\ea~\cite{dean2012ubl} presented an Unsupervised Behavior Learning
framework based on the clustering algorithm. However, cluster-based
approaches do not work well with sparse data, which is the case of our
PNM data where abnormal events are rare. Sung
\ea~\cite{sung2003identifying} deployed Support Vector Machines (SVMs)
to estimate the actual crucial features. According to our evaluation,
SVMs do not perform as well as \sys. Liu \ea~\cite{liu2015opprentice}
adopted more than twenty statistics models to obtain more features
from the original data. They used all generated features in Random
Forest and achieved high accuracy and effectiveness. However, they
still require manual labeling to train the Random Forest model.
PreFix~\cite{zhang2018prefix} predicts switch failures with high
precision and recall. However, it also requires significant manual
efforts for labeling, while our work does not.  Pan
\ea~\cite{pan2017intelligent} also used the tickets as hints to
select potential network faults.  However, they still asked experts to
manually label network faults and use this labelled data to train a
Decision Tree model. In contrast, \sys does not use any manual label.

Previous researches have also focused on processing customer report
tickets. LOTUS~\cite{venkataraman2018assessing} deploys Natural
Language Processing (NLP) techniques to understand the
tickets. Potharaju~\ea~\cite{potharaju2013juggling} built a system
that automatically processes the raw text of tickets to infer the
networking faults and find out the resolution actions.  Jin
\ea~\cite{jin2011making} studied the tickets in cellular networks and
categorized the types of customer trouble tickets. Chen
\ea~\cite{chen2013event} and Hsu \ea~\cite{hsu2011using} use both
customer trouble tickets and social media postings to determine
network outages. This work combines an ISP's customer trouble tickets
and PNM data to infer network faults.

\section{Discussion}

\name uses customer trouble tickets as network fault indicators to
build a classifier without manual labeling. We plan to focus on the
following directions to improve the performance of \name:

\begin{itemize}
    \item When there lacks a large set of labeled data,
      semi-supervised learning~\cite{zhu2009introduction} combines a small set
      of labeled data and a large set of unlabeled data to improve
      classification accuracy. We plan to investigate whether
      semi-supervised learning approach as well as other machine
      learning methods such as deep learning can improve the
      performance of \sys.
    \item Presently, we use network-related tickets to train the
      classifier. We have discovered that customers tend to report
      tickets on weekdays rather than on weekends and during the day
      rather than at night. From this pattern, one may infer that if a
      customer reports a ticket at an ``atypical'' time, it is more
      likely to indicate a customer-impacting problem.  If we place a
      higher weight for such ``outlier'' tickets in a classification
      algorithm, we may increase both the ticket prediction accuracy
      and coverage.

    \item ISPs desire to differentiate failures that affect a group of
      customers from those that affect a single customer.  We refer to
      faults that affect multiple customers as ``maintenance issues.''
      If there is a maintenance issue, it is also desirable to locate
      the place where this issue has happened. It is possible to infer
      maintenance issues by clustering customers' PNM data, and to
      infer the location of a maintenance issue by combining the
      geographical location of each modem with the topology of HFC
      network. It is our future work to study these problems.
      
    \item When detecting network faults, \name outputs whether there
      is an abnormal event and how long it exists. It is desirable to
      rank the severity of abnormal events so that an ISP can
      prioritize its repair actions. It is our future work to
      explore such ranking algorithms.
\end{itemize}

\section{Conclusion}

Cable broadband networks are widely deployed all around U.S. and serve
millions of U.S. households.  However, cable networks have poor
reliability. Although the cable industry has developed a proactive
network maintenance (PNM) platform to address this issue, cable ISPs
have not fully utilized the collected data to proactively detect and
fix network faults.  Existing approaches rely on instantaneous PNM
metrics with manually set thresholds for fault detection and can
introduce an unacceptably high false positive rate.  We design \name,
a system that learns the fault detection criteria from customer
trouble tickets. Our design overcomes the noise from both PNM data and
customer trouble tickets and achieves a nearly four times higher
\precision than the existing tools in the public domain. 
\PENDING{We also employ unsupervised learning on \sys's detection results to automatically diagnose the
type of fault. 
We use a small set of manually labeled data and customer ticket
statistics to evaluate \sys. The evaluation results show that when
compared to the labeled data, \sys achieves a Rand Index of
0.91. During \sys-diagnosed maintenance (or service) events, a much
higher than average frequency of maintenance (or service) tickets
occur, further suggesting that \sys can effectively distinguish
maintenance issues from service issues. The cable ISP we collaborated with has confirmed the effectiveness of \sys with field tests.}

\bibliographystyle{IEEEtran}
\bibliography{paper_ton}

% \bibitem{ref1}
% {\it{Mathematics Into Type}}. American Mathematical Society. [Online]. Available: https://www.ams.org/arc/styleguide/mit-2.pdf

% \end{thebibliography}

\newpage

 

% \end{IEEEbiography}

\vspace{11pt}


\vfill

\end{document}